\documentclass[10pt]{article}
\usepackage{url}
\usepackage{multirow}
\usepackage{amssymb}
\usepackage{graphicx}
\usepackage{color}
\usepackage{hyperref}
\usepackage{url}
\usepackage[normalem]{ulem}
\usepackage{verbatim}
\usepackage{amsmath}
\usepackage{cite}
\usepackage[caption=false,font=footnotesize]{subfig}
\usepackage{multirow}
\usepackage{tikz}
\usetikzlibrary{positioning, shapes}

\DeclareGraphicsExtensions{.pdf,.png,.jpg}
\begin{document}

\newcommand{\eat}[1]{}

\newcommand{\term}[1]{\small {\tt #1}}
\newcommand{\triple}[1]{\small {$\langle${\tt #1}$\rangle$}}

\title{KlusTree: Clustering Answer Trees from Keyword Search on Graphs}
\author{Madhulika Mohanty\\IIT Delhi \and Maya Ramanath\\IIT Delhi}
\date{}

\maketitle
\begin{abstract}

Graph structured data on the web is now massive as well as diverse, ranging from social networks, web graphs to knowledge-bases. Effectively querying this graph structured data is non-trivial and has led to research in a variety of directions -- structured queries, keyword and natural language queries, automatic translation of these queries to structured queries, etc. 
We are concerned with a class of queries called \textit{relationship queries}, which are usually expressed as a set of keywords (each keyword denoting a named entity). The results returned are a set of ranked trees, each of which denotes relationships among the various keywords. The result list could consist of hundreds of answers. The problem of keyword search on graphs has been explored for over a decade now, but an important aspect that is not as extensively studied is that of user experience. We propose KlusTree, which presents clustered results to the users instead of a list of all the results. In our approach, the result trees are represented using language models and are clustered using $JS$ divergence as a distance measure. We compare KlusTree with the well-known approaches based on isomorphism and tree-edit distance based clustering. The user evaluations show that KlusTree outperforms the other two in providing better clustering, thereby enriching user experience, revealing interesting patterns and 
improving result interpretation by the user. 
\end{abstract}
\section{Introduction}
\paragraph*{Motivation and Problem.} Many current, state-of-the-art information systems typically deal with large graphs. These graphs could be entity-relationship graphs extracted from textual sources, relational databases modelled as graphs, biological networks, social networks, or a combination of these. Typically, these graphs are both node-labeled as well as edge-labeled and provide semantic information. Edge weights denote the strength of the relationship between nodes. Since these graphs are often massive, querying and analysing them efficiently is non-trivial. An example of such a massive and fast-growing graph is the Linked Open Data\footnote{\tt http://linkeddata.org} (LOD) graph. 

\begin{figure}[t]
\centering
\subfloat[]{
\begin{tikzpicture}[node distance=1.3cm,on grid]
\node[circle,thick,fill=purple, minimum size=0.2cm,label=Corpse Bride](D){};
\node(P)  [below left of=D,circle,thick,fill=blue, minimum size=0.2cm,label={[align=center]below:Helena\\Carter}] {};
\node(Q)  [below right of=D,circle,thick,fill=blue, minimum size=0.2cm,label={[align=center]below:Johnny\\Depp}] {};
\draw(D) -- (P) node [pos=0.5,sloped,above] {actedIn};
\draw(D) -- (Q) node [pos=0.5,sloped,above] {actedIn};
\end{tikzpicture}}
\subfloat[]{
\begin{tikzpicture}[node distance=1.3cm,on grid]
\node[circle,thick,fill=purple, minimum size=0.2cm,label=Dark Shadows](D){};
\node(P)  [below left of=D,circle,thick,fill=blue, minimum size=0.2cm,label={[align=center]below:Helena\\Carter}] {};
\node(Q)  [below right of=D,circle,thick,fill=blue, minimum size=0.2cm,label={[align=center]below:Johnny\\Depp}] {};
\draw(D) -- (P) node [pos=0.5,sloped,above] {actedIn};
\draw(D) -- (Q) node [pos=0.5,sloped,above] {actedIn};
\end{tikzpicture}}
\subfloat[]{
\begin{tikzpicture}[node distance=1cm,on grid]
\node[circle,thick,fill=purple, minimum size=0.2cm,label=Christopher Lee](D){};
\node(P)  [below left of=D,circle,thick,fill=purple, minimum size=0.2cm,label={[align=right]left:Corpse\\Bride}] {};
\node(Q)  [below right of=D,circle,thick,fill=purple, minimum size=0.2cm,label={[align=left]right:Sleepy\\Hollow}] {};
\node(R)  [below of=P,circle,thick,fill=blue, minimum size=0.2cm,label={[align=center]below:Helena\\Carter}] {};
\node(S)  [below of=Q,circle,thick,fill=blue, minimum size=0.2cm,label={[align=center]below:Johnny\\Depp}] {};
\draw(D) -- (P) node [pos=0.5,sloped,above] {actedIn};
\draw(D) -- (Q) node [pos=0.5,sloped,above] {actedIn};
\draw(P) -- (R) node [midway] {actedIn};
\draw(Q) -- (S) node [midway] {actedIn};
\end{tikzpicture}}\quad
\subfloat[]{
\begin{tikzpicture}[node distance=1.7cm,on grid]
\node[circle,thick,fill=purple, minimum size=0.2cm,label=Tim Burton](D){};
\node(P)  [below left of=D,circle,thick,fill=purple, minimum size=0.2cm,label={[align=right]left:Corpse\\Bride}] {};
\node(Q)  [below right of=D,circle,thick,fill=purple, minimum size=0.2cm,label={[align=left]right:Sleepy\\Hollow}] {};
\node(R)  [below=1cm of P,circle,thick,fill=blue, minimum size=0.2cm,label={[align=center]below:Helena\\Carter}] {};
\node(S)  [below=1cm of Q,circle,thick,fill=blue, minimum size=0.2cm,label={[align=center]below:Johnny\\Depp}] {};
\draw(D) -- (P) node [pos=0.5,sloped,above] {directedBy};
\draw(D) -- (Q) node [pos=0.5,sloped,above] {directedBy};
\draw(P) -- (R) node [midway] {actedIn};
\draw(Q) -- (S) node [midway] {actedIn};
\end{tikzpicture}}
\subfloat[]{
\begin{tikzpicture}[node distance=1.7cm,on grid]
\node[circle,thick,fill=purple, minimum size=0.2cm,label=Walt Disney](D){};
\node(P)  [below left of=D,circle,thick,fill=blue, minimum size=0.2cm,label={[align=right]left:John\\Carter}] {};
\node(Q)  [below right of=D,circle,thick,fill=purple, minimum size=0.2cm,label={[align=left]right:The Lone\\Ranger}] {};
\node(R)  [below=1cm of Q,circle,thick,fill=blue, minimum size=0.2cm,label={[align=center]below:Johnny\\Depp}] {};
\draw(D) -- (P) node [pos=0.5,sloped,above] {producedBy};
\draw(D) -- (Q) node [pos=0.5,sloped,above] {producedBy};
\draw(Q) -- (R) node [midway] {actedIn};
\end{tikzpicture}}
\caption{Results of query \{\emph{``Carter'' ``Depp''}\} over IMDB.}
 \label{fig:SampleQueryResults}
\end{figure}
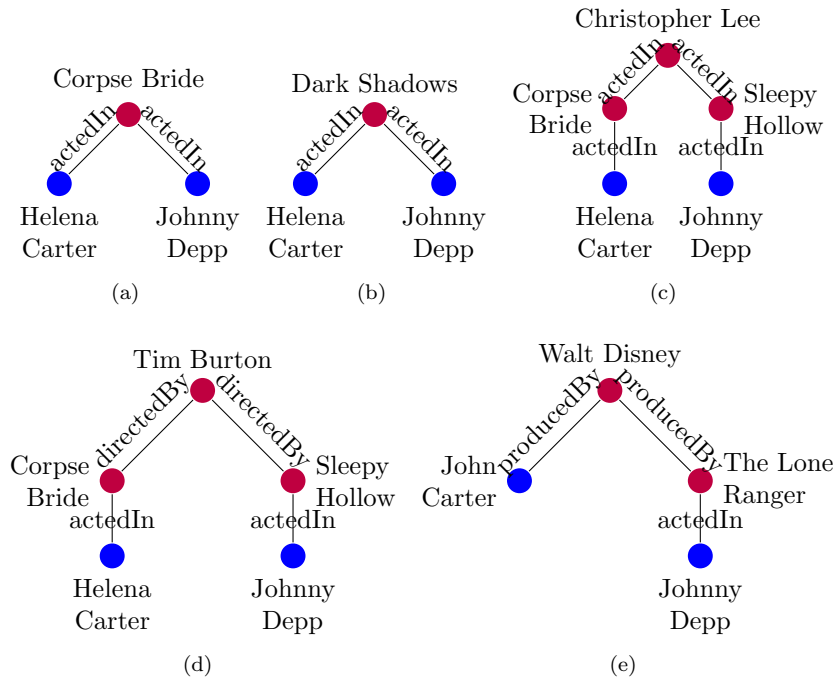
While these graphs can be analysed and queried in a variety of ways, an easy and non-expert way is by using a specific class of queries called \emph{relationship queries}. That is, given a set of two or more entities, what are the relationships that connect all these entities. For example, given entities ``Angela Merkel'', ``Indira Gandhi'', ``Julia Gillard'' and ``Sheikh Hasina'', apart from the interconnection that they are all women, the other (perhaps more interesting) interconnection is that they are/were all heads of their respective governments. Such queries could range from fun and entertainment on graphs of movies and actors to more serious discoveries on graphs of politicians and corporations. In all these cases, the system returns multiple, ranked results \cite{banks1, banks2, DISCOVER1, DISCOVER2, DBXplorer, ObjectRank, xrank, Hristidis@icde2003, Hristidis@TKDE2006}. The results are in the form of trees which express the relationships among the queried entities.  This is 
\textit{keyword search} since the user specifies her query using multiple keywords that denote entities in the graph.

A key problem of keyword search on graphs is that it could potentially return a huge number of results. For example, consider the query consisting of just two keywords, ``Carter'' and ``Depp'' on the IMDB dataset. Example results for this query are shown in Figure \ref{fig:SampleQueryResults}. However, we find that the total number of results returned by this query is 105. While it is certainly possible to restrict the number of results returned to just $k$ (where $k$ is typically 10 or 20), this is counterproductive for the following reasons:

\begin{itemize}
\item If the query is ambiguous, one term could mean two different entities altogether and both the entities could be equally important. Restricting the number of results could result in either missing out one of them or intermixing of results of both types and leading to confusion. For example, for a search query \textit{``Carter''} over IMDB data, a matching node could either be an actor \textit{``Helena Bonham Carter''} or a movie name that has \textit{``Carter''} as its substring, i.e., \textit{``John Carter''}.

\item Displaying only a few answers could also prevent the user from discovering any interesting patterns that might exist in the results. For example, for the query \{\emph{``Carter'',``Depp''}\}, many results are of the form where the two actors, ``Helena Bonham Carter'' and ``Johnny Depp'', have acted in the same movie. Restricting the results may lead to the user missing out on the fact that they have been very frequent coactors.

\end{itemize}

\paragraph*{Approach and Contribution.} In this paper, we propose KlusTree, an approach of clustering over results to avoid the problems highlighted above. KlusTree clusters similar flavours of answers together and thereby allows a single page to encompass a broad variety of results. This helps the users get a bird's eye view of all the different kinds of answers possible and analyse interesting patterns that are now evident.

KlusTree uses a novel, language-model based \cite{ponte1998language} representation of result trees and a standard hierarchical clustering algorithm to cluster the results. In our technique, language models (LMs) are the objects of interest and JS-divergence is a measure of similarity between two LMs. The clusters are ranked using heuristics based on the original ranking of the individual trees given by the underlying keyword search algorithm.
The main advantage of using LMs as opposed to the standard tree similarity measures such as tree edit distance and tree isomorphism is that our LMs can take into account the \emph{neighbourhood} of entities in the result tree. In the example query of \{\emph{``Carter'',``Depp''}\}, even if the given node in the result tree contains just the word ``Carter'', we can still differentiate between the ``Carter''s by looking at the neighbourhood of the node (See Figure \ref{fig:SampleQueryResults}). The node ``Helena Bonham Carter'' would contain the movies she acted in as her neighbours, while the node ``John Carter'' would contain neighbours corresponding to the directors, producers or actors associated with the movie. 

In summary, the main contributions of this paper are as follows:
\begin{enumerate}
 \item A novel language-model based representation of the result trees returned from keyword search over graphs.
 \item A clustering over the answer trees using the language model representation.
 \item User evaluations to compare our approach, KlusTree, with the well-known approaches based on isomorphism and tree-edit distance based clustering, demonstrating the advantage of using the language-model based representation for trees.
 \item Enhanced user experience and result interpretation in keyword search over graphs.
\end{enumerate}

Note that our work is orthogonal to any specific keyword search algorithm and therefore can be used as a post-processing step on any list of results returned by a keyword search algorithm.

\paragraph*{Outline.} The rest of the paper is organized as follows. Section \ref{relatedwork} discusses related work, Section \ref{section:definition} outlines few definitions and Section \ref{mainwork} describes our technique to cluster keyword search results using language model representation of trees. This is followed by Section \ref{experiment} which outlines the evaluation of our technique against the two main baselines and finally, Section \ref{conclude} concludes the paper with a summary of our contributions and future work directions.

\section{Related Work}\label{relatedwork}
We discuss the related work under the following categories:
\subsubsection*{Graph Summarization and Minimization} There has been a lot of work on graph compression and summarization techniques which aim at reducing the graph into a smaller size with minimum loss of information. SNAP\cite{Tian@SIGMOD2008} is an algorithm to approximately describe the structure and content of graph data by grouping nodes based on user-selected node attributes and relationships. This summarization technique exploits the similarities in the non-numerical attributes only. CANAL\cite{Zhang@ICDE2010} extends SNAP to summarize and categorize numerical attributes into various buckets with an option to drill-down or roll-up the resolution of the dataset. \cite{Catherine@COMAD2009} describes an algorithm to generate a summary graph using a random walk model and \cite{Navlakha@SIGMOD2008} compresses graphs using Minimum Description Length(MDL) principle from information theory. All these summary techniques aim at finding a representative graph for the huge data graph. These techniques cannot be 
used in a 
situation where we have multiple result graphs and we are looking for grouping similar results together. Other summarization techniques include those based on bisimulation equivalence relation \cite{Gentilini@JAR2003} and simulation based minimization \cite{Bustan@CADE2000} which summarize graphs by reducing/merging similar nodes and edges.  \cite{Wu@PVLDB2013} uses dominance relation defined in \cite{Henzinger@FOCS1995} to create a summary graph by merging a number of nodes.  
\subsubsection*{Graph Clustering} Graph clustering is the process of grouping many graphs into different clusters such that each cluster contains similar graphs. \cite{Survey@MMGD2010} is an extensive survey on many algorithms based on tree-edit distance for clustering graphs and XML data. All these algorithms are based on tree-edit distance and we have shown a comparison of our technique with tree-edit distance based technique described in \cite{tree-edit}.

\paragraph*{Tree-edit distance based clustering}
This approach has been mainly used for clustering XML documents that rely on a tree representation of the documents. The trees can be converted into documents by having the nodes and relationships labelled by their textual content. The problem of computing the distance between two trees, also known as tree editing problem, is the generalization of the problem of computing the distance between two strings to labelled trees \cite{TreeEditKlein}. The editing operations available in the tree editing problem are changing (i.e., relabelling), deleting, and inserting a node/relationship. To each of these operations a cost is assigned. The edit distance problem is to find a sequence of such operations transforming a tree $T_{1}$ into a tree $T_{2}$ with minimum cost. The distance between $T_{1}$ and $T_{2}$ i.e $TED(T_{1},T_{2})$ is then defined to be the cost of such a sequence. We have shown a comparison of our approach with a tree-edit distance based clustering approach. The basic steps in this approach have been 
adopted from \cite{tree-edit} and are as follows. Let us consider the set $\{T_{1},...,T_{n}\}$ of all the trees.
\begin{enumerate}
\item Compute the similarity matrix($M$). This is a $n \times n$ matrix where the $(i,j)$ position (denoted $M_{i,j}$) is obtained as $es(T_{i}, T_{j})$, the edit-distance similarity between $T_{i}$ and $T_{j}$, where 
\begin{equation}
es(T_{i}, T_{j}) = 1 - \dfrac{TED(T_{i}, T_{j})}{nodes(T_{i})+ nodes(T_{j})}.
\end{equation}
The function $nodes(T)$ returns the number of nodes in tree $T$. $es(T_{i}, T_{j})$ is a similarity value for $T_{i}$ and $T_{j}$ between $0$ and $1$.
\item The column similarity between $T_{i}$ and $T_{j}$, denoted $cs(T_{i}, T_{j})$, is defined as the inverse of the average absolute error between the columns corresponding to $T_{i}$ and $T_{j}$ in the
similarity matrix. 
\begin{equation}
cs(T_{i},T_{j})= 1 - \sum_{k=1.n} \arrowvert M_{i,k}-M_{j,k} \arrowvert / n
\end{equation}

Therefore, to consider two subtrees as similar, the column similarity measure requires their columns in the similarity matrix to be very similar. This means two subtrees must have roughly the same edit-distance similarity with respect to the rest of subtrees in the set to be considered as similar. Column similarity has been found to be more robust for estimating similarity between $T_{i}$ and $T_{j}$ in the clustering process than directly using $es(T_{i}, T_{j})$.
\item Now, using these values as a distance measure, any clustering algorithm can be used to cluster the trees. We have used the same algorithm as the one used for our approach.
\end{enumerate}

\subsubsection*{Text/XML Summarization} These techniques summarize textual documents and XML trees. Grouper \cite{Grouper} and \cite{Anastasiu@CIKM2011} propose techniques to cluster textual document results from a keyword search on unstructured data. Tag clouds are another very popular way to summarize web search results \cite{Kuo@WWW2007}. Emails can be summarized using summary keywords generated using Latent Sematic Analysis (LSA) and Latent Dirichlet Allocation (LDA) \cite{Dredze@IUI2008}. These techniques for clustering and summarizing text cannot be directly applied to trees. \cite{Huang@SIGMOD2008} generates summaries for XML results (as trees), where a snippet is produced for each result tree. The snippets with similar structures can be grouped together for better understanding \cite{Liu@IDEB2010}. Snippet generation does not take into account the grouping together of similar results. These techniques can be applied over a clustering technique to generate summaries for each cluster. 

Many isomorphism based result clustering algorithms have been proposed for clustering XML results. TreeCluster \cite{TreeCluster} uses two-level summarization to summarize results. It first divides the results into groups using graph isomorphism. It further divides them on the basis of keyword frequencies. Such techniques are limited by their isomorphism based approach which takes only high structural and content similarity into account and does not consider context. Other approaches include XSeek\cite{Liu@ATDS2010} which uses a search keywords' ancestor based splitting to divide the results into $k$ clusters and \cite{Gkorgkas@ADBIS2013} which uses cosine similarity between the tf-idf vectors of results as a similarity measure to cluster the trees. . 

\subsubsection*{Keyword Search Result Enhancement} This class of algorithms aims at refining a keyword query for getting better results. \cite{Koutrika@EDBT2009} describes algorithms to find and show good and most relevant keywords from the results, which can further be used for query refinement. Faceted Search \cite{Ben-Yitzhak@WSDM2008, Roy@CIKM2008} aims at providing query refinement in the form of a navigational search. It starts by providing results classified into different categories. The user gets more refined results when she selects one of the categories. Given a set of clustered results, \cite{Liu@PVLDB2011} suggests a query expansion scheme based on F-measure. These techniques aim at improving the result quality by improving upon the preliminary results itself but they do not address the issue of easy comprehension and understandability of the results.

\subsubsection*{Coverage aware approaches} The coverage aware search techniques aim at finding an answer with the most diverse nature so as to have a high probability of satisfying the user. DisC\cite{Drosou@PVLDB2012} gives a concept of DisC diversity and proposes a greedy algorithm for finding diverse results. Other greedy based techniques are given in \cite{Agrawal@WSDM2009}, \cite{Ranu@SIGMOD2014} and \cite{Angel@SIGMOD2011}. These techniques can be used to find the cluster representatives. We have used a simple way of picking up the best ranked result as the representative of a cluster but these concepts can be adapted to get better representatives. Since currently our main focus is on finding good clusters and not on finding a good representative, we do not compare our results with these techniques.

\section{Definitions} \label{section:definition}
We first lay out few definitions that will be used henceforth.

\paragraph*{\textbf{Graph}} Any graph structured data is represented by a directed labelled graph, represented as $G = (V, E)$ where $V$ is the vertex set and $E$ is the edge set. In the graph structured data, each directed link from a subject to an object is considered to be a triple of the form, $\langle S,P,O\rangle$ where $S$ is the subject node, $P$ is the edge annotation and $O$ is the object node.  

\paragraph*{\textbf{Keyword Search}} One of the means to query graphs is to use ``keyword'' queries. It consists of $n$ keywords, $Q=\{k_{1},....,k_{n}\}$, where $k_{i}$ refers to the $i$\textsuperscript{th} keyword. 

\paragraph*{\textbf{Answer}} Each result denotes a ``relationship'' between the query keywords and is in the form of a directed subtree $T$ of the data graph $G$, such that $T$ contains all the keywords of $Q$ and is minimal. Thus, the subtree $T$ must include the nodes for all the keywords of $Q$. Minimality of $T$ means that $T$ is minimal with respect to $Q$, i.e., $T$ contains the nodes for all the keywords of $Q$, but has no proper directed subtree that also includes all of them. The root node $r$ is such that every other node of $T$ is reachable from $r$ through a unique directed path. 

Many algorithms have been proposed for doing a keyword search on graph structured data. Few notable ones are BANKS\cite{banks1}, Bidirectional \cite{banks2}, DISCOVER \cite{DISCOVER1,DISCOVER2}, DBXplorer \cite{DBXplorer} and ObjectRank \cite{ObjectRank}. Some algorithms for keyword search on graphs also consider graphs as results \cite{EASE@SIGMOD2008}. Though we consider trees as results, our technique can easily be applied to graphs as well without any significant changes.

\section{Clustering Keyword Search Results}\label{mainwork}

\begin{table}[!h]
    \centering   
    \subfloat[Document of Entity \textit{Corpse Bride}]{
\begin{tabular}{ |p{4cm}| p{2cm} |p{2.5cm}| }
\hline\multicolumn{3}{|c|}{\textbf{Corpse Bride}}\\

\hline
Corpse Bride & DirectedBy & Tim Burton \\ 
Helena Carter & ActedIn & Corpse Bride \\ 
Johnny Depp & ActedIn & Corpse Bride \\ 
Corpse Bride & hasLanguage & English \\ 
\hline
\end{tabular} \label{tab:DocOfEnta}}\\
\subfloat[Document of Entity \textit{Johnny Depp}]{
\begin{tabular}{  |p{2.5cm}| p{2cm}| p{4cm}| }
\hline
\multicolumn{3}{|c|}{\textbf{Johnny Depp}} \\
\hline
Johnny Depp & ActedIn & Corpse Bride \\ 
Johnny Depp & ActedIn & Sleepy Hollow \\ 
Johnny Depp & ActedIn & The Lone Ranger \\ 
\hline
\end{tabular}\label{tab:DocOfEntb} }\\
\subfloat[Unigrams and Bigrams in the Document for entity \textit{Corpse Bride}]{
\begin{tabular}{|p{1.5cm}| p{9cm}|}
\hline 
\multicolumn{2}{|c|}{\textbf{Corpse Bride}}\\
\hline
 Unigrams & \{Tim Burton, Helena Carter, Johnny Depp, English\} \\ 
\hline
 Bigrams & \{(DirectedBy, Tim Burton), (Helena Carter, ActedIn), (Johnny Depp, ActedIn), (hasLanguage,English)\} \\
\hline
\end{tabular}\label{tab:UnBa}} \\
\subfloat[Unigrams and Bigrams in the Document for entity \textit{Johnny Depp}]{
\begin{tabular}{|p{1.5cm}| p{9cm}|}
\hline
\multicolumn{2}{|c|}{\textbf{Johnny Depp}} \\
\hline
 Unigrams & \{Corpse Bride, Sleepy Hollow, The Lone Ranger\} \\ 
\hline
 Bigrams & \{(ActedIn, Corpse Bride), (ActedIn, Sleepy Hollow), (ActedIn, The Lone Ranger)\} \\
\hline
\end{tabular}\label{tab:UnBb} } \\
\subfloat[Final LMs For Entities \textit{Corpse Bride} and \textit{Johnny Depp}]{
    
 \begin{tabular}{|p{1.5cm} | p{1.5cm} | p{1.5cm} | p{1cm} | p{2.5cm} |p{2cm} |p{1cm}|  }
\hline
\textbf{Entity} & \textbf{Helena Carter} & \textbf{Sleepy Hollow} & \textbf{...} & \textbf{(hasLanguage, English)} & \textbf{(ActedIn, Sleepy Hollow)}& \textbf{...}\\ \hline
Corpse Bride & 0.08 & 0.02 & ... & 0.09 & 0.02 &...\\ 
Johnny Depp &  0.06 & 0.08 & ... & 0.03 & 0.07&... \\ 

\hline
\end{tabular}\label{tab:FinalEntLM}}
\caption{Table showing LM for entities}
    
\end{table}
Our paper proposes a technique, KlusTree, to cluster the answers given out by the keyword search algorithms. The clustering is based on language models (LMs) \cite{ponte1998language}. 

\subsubsection*{Language Models} A statistical language model assigns a probability to a sequence of $m$ words, $P(w_{1},...,w_{m})$, by means of a probability distribution using the maximum likelihood estimate and smoothes using a background model. Let $w$ be a word in the document $D$. Then, its probability in the document model is given by:
\begin{equation}
 P(w)=\lambda P(w|D) + (1-\lambda)P(w|C)
\end{equation}
where $P(w|D)$ is the probability of $w$ in document $D$, $P(w|C)$ is the probability of $w$ in the entire corpus $C$ and $\lambda$ is the smoothing parameter.\\

We first estimate the models of entities and relationships in the graph. Based on these, we estimate the language model of each result tree, and use it to cluster the results. We adopt the approach introduced in \cite{ElbassuoniRW11} for estimating the language models.

\subsection{LMs of Entities}
\subsubsection*{Document of an entity}
Language models are estimated for a \emph{document} over a set of \emph{terms}. Therefore, estimating an LM for an \emph{entity}, first requires us to define a ``pseudo-document'' for an entity and then terms in this document. We do this as follows. A document $D$ of an entity $\mathbf{E}$ is a set of all triples from our dataset where this entity occurs.
\begin{equation}
D(\mathbf{E}) = \{{\langle \mathbf{E} P O \rangle:\langle \mathbf{E} P O \rangle \in G}\} \cup \{{\langle S P \mathbf{E} \rangle:\langle S P \mathbf{E} \rangle \in G}\}
\end{equation}
where $S$ is the subject, $P$ is the predicate, $O$ is the object and $G$ is the data graph. Table \ref{tab:DocOfEnta} and \ref{tab:DocOfEntb} show the documents corresponding to the entities \textit{Corpse Bride} and \textit{Johnny Depp}.

We define two types of terms, unigrams and bigrams, as follows. Unigrams are the set of objects occurring in triples where $\mathbf{E}$ is a subject and the set of subjects occurring in triples where $\mathbf{E}$ is an object, i.e.,  $U(\mathbf{E}) = \{{O :\langle \mathbf{E} P O \rangle \in G}\} \cup \{{S :\langle S P \mathbf{E} \rangle \in G}\}$. Similarly, bigrams are defined as: $B(\mathbf{E}) = \{{P O :\langle \mathbf{E} P O \rangle \in G}\} \cup \{{S P :\langle S P \mathbf{E} \rangle \in G}\}$. Table \ref{tab:UnBa} and \ref{tab:UnBb} show the set of unigrams and bigrams for the entities \textit{Corpse Bride} and \textit{Johnny Depp}.

\subsubsection*{Estimating LM of entities}
The LM corresponding to document $D(\mathbf{E})$ for an entity $\mathbf{E}$ is a mixture model of two LMs: $P_{U(\mathbf{E})}$, corresponding to the unigram LM for $\mathbf{E}$ and $P_{B(\mathbf{E})}$, corresponding to the bigram LM for $\mathbf{E}$. That is,
\begin{equation}
P_{\mathbf{E}}(w) = \mu P_{U(\mathbf{E})}(w) + (1 - \mu)P_{B(\mathbf{E})}(w) 
\end{equation}
where $w$ is a term over which the language model of $\mathbf{E}$ had to be calculated and $\mu$ controls the influence of each component (equal weights were assigned for both the components during experiments). Both unigram and bigram LMs are estimated in the standard way using maximum likelihood estimate and smoothing from the data corpus. Let $w$ be a term in the given query. Then, the probability of $w$ in the document unigram language model is given by:
\begin{equation}
P_{U(\mathbf{E})}(w) = \lambda \frac{c(w, D(\mathbf{E}))}{\sum_{w' \in U(\mathbf{E})}c(w',D(\mathbf{E}))} + (1 - \lambda) \frac{c(w,D(G))}{\sum_{w' \in U(\mathbf{E})}c(w', D(G))}
\end{equation}
where $c(w, D(\mathbf{E}))$ is the count of word $w$ in the document of entity $\mathbf{E}$ and $\sum_{w' \in U(\mathbf{E})}c(w',D(\mathbf{E}))$ is the count of all the unigram terms of $E$ in the document of entity $\mathbf{E}$. So the first term in above equation is the probability of $w$ in the document with respect to only unigram terms. Similarly, $c(w, D(G))$ is the count of word $w$ in the Data Graph and
$\sum_{w'\in U(\mathbf{E})}c(w',D(G))$ is the count of all terms of unigrams of $\mathbf{E}$ in the Data Graph. So the second term in the above equation is the probability of $w$ in the corpus, $G$ with respect to only unigram terms. $\lambda$ is the smoothing parameter and was set to $0.5$. A similar approach was used to compute Bigram LMs.

Table \ref{tab:FinalEntLM} shows the final picture of LMs for \textit{Corpse Bride} and \textit{Johnny Depp}.

\subsection{LMs of Relationships}
\subsubsection*{Document of a relationship}
We build a document for each relationship in the dataset. A document $D$ of a relationship $\mathbf{R}$ is a set of all triples from our dataset where this relationship occurs i.e. $D(\mathbf{R}) = \{{\langle S \mathbf{R} O \rangle:\langle S \mathbf{R} O \rangle \in G}\}$. There are two types of terms over each relationship $\mathbf{R}$, unigrams and bigrams. Subject unigrams are  $S(\mathbf{R}) = \{{S :\langle S \mathbf{R} O \rangle \in G}\}$, object unigrams are $O(\mathbf{R}) = \{{O :\langle S \mathbf{R} O \rangle \in G}\}$ and bigrams are $B(\mathbf{R}) = \{{S O :\langle S \mathbf{R} O \rangle \in G}\}$. Tables \ref{tab:RelationDoc} shows the document of the relationship \textit{ActedIn} and Table \ref{tab:termsRel} shows the terms to be considered for the LM.

\begin{table}[!t]
    \centering   
\begin{tabular}{ |p{4cm}| p{2cm}| p{4cm}| }
\hline
\multicolumn{3}{|c|}{\textbf{ActedIn}}\\
\hline
Johnny Depp & ActedIn & Sleepy Hollow \\ 
Helena Carter & ActedIn & Corpse Bride \\ 
Johnny Depp & ActedIn & The Lone Ranger \\ 
\hline
\end{tabular} 
    \caption{Document of Relationship \textit{ActedIn}}
    \label{tab:RelationDoc}
\end{table}

\begin{table}[!h]
     \centering
\begin{tabular}{|p{2.5cm}| p{9cm}|}
\hline
 \multicolumn{2}{|c|}{\textbf{ActedIn} }\\
\hline
 Subject unigrams & \{Johnny Depp,Helena Carter\} \\ 
\hline
 Object unigrams & \{Corpse Bride, Sleepy Hollow, The Lone Ranger\} \\ 
\hline
 Bigrams & \{(Johnny Depp, Sleepy Hollow), (Johnny Depp, The Lone Ranger), (Helena Carter, Corpse Bride)\} \\
\hline
\end{tabular} 
    \caption{Terms in the document of relationship \textit{ActedIn}}
    \label{tab:termsRel}
\end{table}

\subsubsection*{Estimating LM of relationships}
The LM corresponding to document $D(\mathbf{R})$ for a relationship $\mathbf{R}$ is a mixture model of three LMs: $P_{S(\mathbf{R})}$, corresponding to the subject unigram LM of $\mathbf{R}$, $P_{O(\mathbf{R})}$, corresponding to the object unigram of $\mathbf{R}$ and $P_{B(\mathbf{R})}$, corresponding to the bigram LM of $\mathbf{R}$. That is,
\begin{equation}
P_{\mathbf{R}}(w) = \mu_{s}P_{S(\mathbf{R})}(w) + \mu_{o}P_{O(\mathbf{R})}(w) + (1 - \mu_{s} - \mu_{o})P_{B(\mathbf{R})}(w) 
\end{equation}
where $\mu_{o}$ and $\mu_{s}$ control the influence of each component. We have given equal weights to each component during our evaluations. The unigram and bigram LMs are estimated in the same way as the LMs of entities are calculated. 

That is,

\begin{equation}
P_{S(\mathbf{R})}(w) = \lambda \frac{c(w, D(\mathbf{R}))}{\sum_{w' \in S(\mathbf{R})}c(w',D(\mathbf{R}))} + (1 - \lambda) \frac{c(w,D(G))}{\sum_{w' \in S(\mathbf{R})}c(w', D(G))}
\end{equation}
\begin{equation}
P_{O(\mathbf{R})}(w) = \lambda \frac{c(w, D(\mathbf{R}))}{\sum_{w' \in O(\mathbf{R})}c(w',D(\mathbf{R}))} + (1 - \lambda) \frac{c(w,D(G))}{\sum_{w' \in O(\mathbf{R})}c(w', D(G))}
\end{equation}
\begin{equation}
P_{B(\mathbf{R})}(w) = \lambda \frac{c(w, D(\mathbf{R}))}{\sum_{w' \in B(\mathbf{R})}c(w',D(\mathbf{R}))} + (1 - \lambda) \frac{c(w,D(G))}{\sum_{w' \in B(\mathbf{R})}c(w', D(G))}
\end{equation}

\subsection{LMs of Result Trees} 
As described earlier, a result tree is a rooted tree having entities as nodes and relationships as edges. The LM of the result tree comprises of two kinds of LMs, entity LM and relationship LM. The entity LM captures the content information, while the relationship LM represents the structural information.

We use a mixture model of component entity LMs to estimate the entity LM for each result tree. We first get the LM vector of all the entities present in the tree and then we mix them term-wise to estimate the entity LM for a result tree.  The Entity LM of a tree $T$ from the result trees having entities $E_{1}$,$E_{2}$,... is given by:

\begin{equation}
 P_{E}(T)=(\delta_{1}*P_{E_{1}}) + (\delta_{2}*P_{E_{2}})+ ...
 \label{eq:TreeLM}
\end{equation}
where $\delta_{1}$,$\delta_{2}$,... refer to the weights of the respective entities.

Entity Vector of a tree is the term-wise weighted sum of LMs of all the entities present in the trees. We have given equal weight to all the entities in our calculations and hence all the $\delta$s are equal and sum to 1. We can also learn the weights using various learning techniques, but assigning equal weights has been found to work in practice.

The tree's relationship LM is estimated in a similar way.

\subsection{Clustering Result Trees}
Once the LMs of result trees are calculated, they can be clustered using a clustering algorithm. 
\paragraph*{\textbf{Similarity Measure}}
We use $JS$-Divergence as the similarity measure between the tree LMs. Divergence between two trees is calculated by weighted sum of $JS$-Divergence of entity and relationship LMs of the trees.
\begin{equation}
JS(T_{1}||T_{2}) = \gamma ∗ JS_{E}(T_{1}||T_{2}) + (1 - \gamma) ∗ JS_{R}(T_{1}||T_{2}) 
\end{equation}
where $\gamma$ controls the weights we want to give to entity LM and relationship LM. We have given equal weights to both the LMs and hence, $\gamma = 0.5$ in our experiments.

\paragraph*{\textbf{Clustering Algorithm}}

We have used the hierarchical complete-link clustering algorithm. It is a method of cluster analysis which seeks to build a hierarchy of clusters. We have adopted Agglomerative Complete-Link Clustering technique. Agglomerative strategy is a ``bottom up'' approach: each observation starts in its own cluster, and pairs of clusters are merged as one moves up the hierarchy. Complete-Link Clustering technique considers the similarity of two clusters as the similarity of their most dissimilar members. This is equivalent to choosing the cluster pair whose merge has the smallest diameter. This tends to create more tightly centered clusters. The optimum number of clusters is determined by using the Calinski-Harabasz (CH) index \cite{calinski1974}. For a given number of trees $N$ and for a range of values for no. of clusters, $K$, the one with the highest CH index value is chosen as the best clustering.

 For a given number of trees $N$ and clusters, $K$, it is computed as follows:
\begin{equation}
 W(K) = \frac{1}{2}\sum\limits_{k=1}^K \sum_{T_{i} \in k^{th} Cluster} \sum_{T_{j} \in k^{th} Cluster} JS(T_{i}||T_{j})
 \end{equation}
 \begin{equation}
 B(K) = \frac{1}{2}\sum\limits_{k=1}^K \sum_{T_{i} \in k^{th} Cluster} \sum_{T_{j} \notin k^{th} Cluster} JS(T_{i}||T_{j})
 \end{equation}
 \begin{equation}
 CH(K) = \frac{(N-K) * B(K)}{(K-1) * W(K)}
\end{equation}
where $W(K)$ is the within-cluster scatter, $B(K)$ is the between-cluster scatter and $CH(K)$ is the CH index value for $K$ clusters.

\subsection{Ranking clusters}\label{subsec:rankingdesc}

Result trees returned by the keyword search implementation are ranked. These ranks can be used to get a ranking on clusters. Four variations of ranking have been used:
\begin{enumerate}
 \item Best ranked tree based: The best ranked tree is picked up from each cluster and the clusters are then ranked on the basis of these representatives.
 \item Worst ranked tree based: The worst ranked tree is picked up from each cluster. The clusters are then ranked on the basis of these trees. The cluster having a good worst ranked tree is ranked higher than others.
 \item Average rank of the trees based: The average of all the trees' rank is computed and the clusters are ranked on the basis of it.
 \item Size of the trees based: The largest tree size is computed for each cluster. The cluster having largest amongst the largest trees is ranked lowest.
\end{enumerate}

\section{Experiments}\label{experiment}
In this section, we summarize our experimental findings with respect to the following: (a) the performance of our method compared to tree-edit distance based and graph isomorphism based clustering and, (b) the quality of ranking achieved by the various heuristics.
\subsection{Setup}
\paragraph*{\textbf{Dataset}}
We have used the IMDB dataset given on the website of IMDB\footnote{\url{http://www.imdb.com/interfaces}}. When represented as a graph, it has about 2 million nodes and 9 million edges.

\paragraph*{\textbf{Keyword Queries}}
We manually chose 15 queries as our benchmark. The no. of keywords in the queries range from $2$ to $4$. Each keyword corresponds to an entity. The queries are formed by mixing different \emph{types} of entities. We constructed these queries manually (rather than generate them automatically) to make sure that the results would be interpretable by users. 
\paragraph*{\textbf{Competitors}}
Our baseline is a clustering using a simple distance measure based on tree-isomorphism where two trees are similar only if their \emph{structures} are identical (no content in the form of node and edge labels are considered). Our other competitor is a tree-edit distance measure \cite{tree-edit}. This measure takes into account both the structure as well as the content of the trees.
\paragraph*{\textbf{Methodology and Metrics}}
We generated the set of answer trees for each keyword query using the Bidirectional algorithm \cite{banks2}. Out of all the trees given as output, we picked up the top $25$ trees since the later results became increasingly complex to interpret. In the case of isomorphism, two trees were clustered together iff they were isomorphic to each other. In the case of tree-edit distance and LM based approaches, we used agglomerative complete-link clustering for constructing the dendogram. The optimum number of clusters ($K$) was determined by using $CH$ index. We chose a range of $K$ from $0$ to $15$. The optimum $K$ was decided by choosing the one with highest $CH$ index value. The clusters generated were ranked using each of the heuristics: Best, Worst, Average and Largest tree based. \\
We evaluated the following:
\begin{enumerate}
 \item Clustering Quality: 
 We need to evaluate the performance of our LM-based distance measure as opposed to isomorphism and tree-edit distance. In order to do this, we asked $15$ volunteers to evaluate the quality of the clusters generated. We chose the maximum and minimum similarity (distance) pairs within each cluster generated by the three algorithms (Only one pair per cluster for isomorphism since within a cluster, all the trees are isomorphic to each other). We also picked the representatives of each cluster (the one with the highest rank within the cluster) and constructed pairs for each combination of two clusters. The pairs so generated were shown to the users. They rated each pair on a scale of 1 (Highly Different) to 5 (Highly Similar) on the basis of whether they thought the interpretation of the results were similar or not. The pairs were shown randomly and the users were unaware of the algorithm used and whether these belonged to the same cluster or different clusters. The users were given instructions to first 
interpret the results and then judge whether they conveyed similar information or not. The scores given were then used to judge the semantic similarity within clusters and semantic separation between clusters. Higher the value of the score for pairs within the same cluster, the better is the similarity between the trees within a cluster. Lower the value is for the pairs across clusters, the better is the separation between clusters.
 \item Cluster Ranking Quality:
 For ranking evaluations, the users were shown cluster representatives for each cluster for every query. They were asked to give a score based on the \textit{relevance} of the result, given the query, on a scale of 1 (Very uninteresting) to 5 (Highly Interesting). We then used these scores to compute the Normalized Discounted Cumulative Gain (NDCG) value \cite{NDCG} and judged the ranking techniques.
\end{enumerate}

Therefore, we evaluate, semantic similarity within clusters, semantic separation between clusters and cluster ranking.

\subsection{Results}

\subsubsection{Clustering Quality}
\begin{table}[!t]
    \centering   
		\begin{tabular}{|p{3cm} | p{3.5cm} | p{3.5cm}|}
\hline			
			\textbf{Technique} & Cluster Cohesion (higher the better) & Cluster Separation (lower the better) \\ 
\hline
Isomorphism & 4.0& 2.6\\
LM & 4.2& 2.4\\
Tree-Edit Distance & 3& 2.9\\
\hline
\end{tabular}
\caption{Average values of the evaluation metrics for the 3 techniques.}
\label{tab:results}
\end{table}

The average values for the similarity for the three techniques are summarized in Table \ref{tab:results}. The results are described below:

\paragraph*{Semantic Similarity within clusters}
	
\begin{figure}[t]
\centering
\subfloat[]{
\begin{tikzpicture}[node distance=1cm,on grid]
\node[circle,thick,fill=purple, minimum size=0.2cm,label=David Hyde Pierce](D){};
\node(P)  [below left of=D,circle,thick,fill=purple, minimum size=0.2cm,label={[align=right]left:Full\\Frontal}] {};
\node(Q)  [below right of=D,circle,thick,fill=purple, minimum size=0.2cm,label={[align=left]right:56th Annual Primetime\\Emmy Awards}] {};
\node(R)  [below of=P,circle,thick,fill=blue, minimum size=0.2cm,label={[align=center]below:David\\Fincher}] {};
\node(S)  [below of=Q,circle,thick,fill=blue, minimum size=0.2cm,label={[align=center]below:Brad\\Pitt}] {};
\draw(D) -- (P) node [pos=0.5,sloped,above] {actedIn};
\draw(D) -- (Q) node [pos=0.5,sloped,above] {actedIn};
\draw(P) -- (R) node [midway] {actedIn};
\draw(Q) -- (S) node [midway] {actedIn};
\end{tikzpicture}\label{fig:IsovsLMa}}\quad
\subfloat[]{
\begin{tikzpicture}[node distance=1cm,on grid]
\node[circle,thick,fill=purple, minimum size=0.2cm,label=Gregory Sporleder](D){};
\node(P)  [below left of=D,circle,thick,fill=purple, minimum size=0.2cm,label={[align=right]left:Being John\\Malkovich}] {};
\node(Q)  [below right of=D,circle,thick,fill=purple, minimum size=0.2cm,label={[align=left]right:True\\Romance}] {};
\node(R)  [below of=P,circle,thick,fill=blue, minimum size=0.2cm,label={[align=center]below:David\\Fincher}] {};
\node(S)  [below of=Q,circle,thick,fill=blue, minimum size=0.2cm,label={[align=center]below:Brad\\Pitt}] {};
\draw(D) -- (P) node [pos=0.5,sloped,above] {actedIn};
\draw(D) -- (Q) node [pos=0.5,sloped,above] {actedIn};
\draw(P) -- (R) node [midway] {actedIn};
\draw(Q) -- (S) node [midway] {actedIn};
\end{tikzpicture}}
\subfloat[]{
\begin{tikzpicture}[node distance=1cm,on grid]
\node[circle,thick,fill=purple, minimum size=0.2cm,label=Willie Garson](D){};
\node(P)  [below left of=D,circle,thick,fill=purple, minimum size=0.2cm,label={[align=right]left:Being John\\Malkovich}] {};
\node(Q)  [below right of=D,circle,thick,fill=purple, minimum size=0.2cm,label={[align=left]right:Across}] {};
\node(R)  [below of=P,circle,thick,fill=blue, minimum size=0.2cm,label={[align=center]below:David\\Fincher}] {};
\node(S)  [below of=Q,circle,thick,fill=blue, minimum size=0.2cm,label={[align=center]below:Brad\\Pitt}] {};
\draw(D) -- (P) node [pos=0.5,sloped,above] {actedIn};
\draw(D) -- (Q) node [pos=0.5,sloped,above] {actedIn};
\draw(P) -- (R) node [midway] {actedIn};
\draw(Q) -- (S) node [midway] {actedIn};
\end{tikzpicture}}
\caption{Difference between clusterings by isomorphism and LMs for the query - ``Brad Pitt'' ``David Fincher''.}
 \label{fig:IsovsLM}
\end{figure}
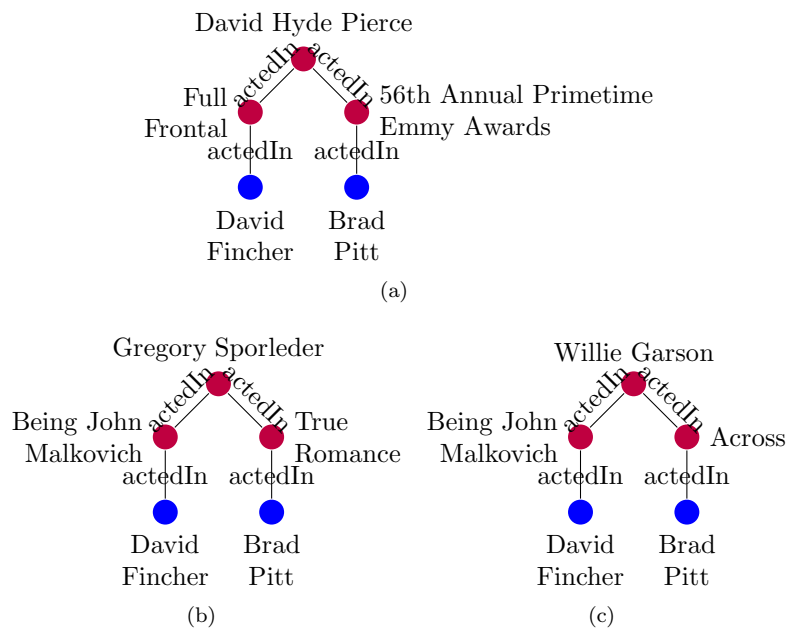

The average cluster cohesion score for isomorphism based approach over all queries comes out to be 4.0/5 which implies trees within the same cluster were conveying \emph{Similar} information upto a large extent. It shows that isomorphism performs well on this criteria. It is also very obvious because trees should have exactly same structure, if they are to be clustered together. Hence, cluster cohesion is very high in this approach. The average cluster cohesion score for tree-edit distance based approach over all queries comes out to be 3/5, which is the lowest amongst all approaches. This is because tree-edit distance does not consider the context at all. In most of the cases, the information conveyed changed completely on adding/deleting few edges and nodes of a tree to arrive at the other tree but the tree-edit distance approach clustered these trees together since the edit-distance was low. The average cluster cohesion score for LM based approach over all queries comes out to be 4.2/5, which is the 
highest among all approaches and is very close to that of isomorphism. This shows that LM based approach led to clusters which were found to convey \textit{highly similar} information. It is evident that LM based approach's performance is comparable to other approaches on this criteria. Clustered trees had both content and semantic similarity. One of the cases where LMs performed better than Isomorphism is shown in Figure \ref{fig:IsovsLM}. The isomorphism based approach clusters all the three trees together whereas the LM based approach puts the tree in Figure \ref{fig:IsovsLMa} into a separate cluster as it has a node which represents an ``Award''.

\paragraph*{Semantic Separation between clusters}
Isomorphism and LMs perform very good on this criteria since the trees in different clusters were found to convey entirely different pieces of information. The average score for isomorphism based approach for all queries comes out to be 2.6/5 which is between \emph{Different and Neutral similarity}, which implies good separation between the information provided by two different clusters. The average of similarity scores for tree-edit distance based approach over all queries comes out to be 2.9/5 which is closer to \emph{Neutral}. This shows that this technique performs poorly on this criteria. It was unable to maintain high separation between different clusters. The average of scores for LM based approach over all queries comes out to be 2.4/5 which is between \emph{Different and Neutral similarity}. This means that the users found two trees from different clusters to be conveying different information. Also, LM based approach outperforms the other two approaches on this criteria. The reason is that our 
approach considers both the content similarity by incorporating Entity LM vectors and structural similarity by incorporating Relationship LM vectors. Hence, if results are placed in different groups they tend to be highly dissimilar. This shows that LM based approach performs very well in separating information provided by two clusters.

\subsubsection{Cluster Ranking Quality}
We have used the NDCG metric to evaluate our ranking technique. It gives a value between $0$ and $1$ with $1$ being the ideal ranking. The average NDCG values for each type of ranking are given in Table \ref{tab:resultsranking}. These results show that the ranking mechanism performs well and is sufficient for the ranking of the clusters. Also, all the techniques perform equally good with the best being the one based on using best ranked tree of the clusters.

\begin{table}[!htbp]
    \centering   
		\begin{tabular}{|p{4.5cm} | p{3cm} |}
\hline			
			\textbf{Technique} & Average NDCG  \\ 
\hline
Best ranked tree based & 0.97\\
Worst ranked tree based & 0.96\\
Average rank of the trees based & 0.95\\
Size of the trees based & 0.96\\
\hline
\end{tabular}
\caption{Average values of the NDCG for the four ranking techniques.}
\label{tab:resultsranking}
\end{table}

\section{Conclusions}\label{conclude}
In this paper, we have proposed KlusTree, a language model based clustering for the results of keyword search on a graph structured data. By providing a clustering of these results, we can improve user experience, result understanding and also reveal interesting patterns. We have also conducted user evaluations on the clusters generated using KlusTree against the two well-known techniques, i.e., isomorphism and tree-edit distance based clustering. These evaluations verified that our technique performs very well in grouping similar results together and thereby, enhances result interpretation and enriches user experience. For future work, we would like to design a good technique to generate summary snippets for each cluster. We would also like to find a good representative for a cluster rather than just picking up the best ranked tree.

\end{document}